\documentclass[conference]{IEEEtran}
\IEEEoverridecommandlockouts
\usepackage{cite}
\usepackage{amsmath,amssymb,amsfonts}
\usepackage{algorithmic}
\usepackage[ruled,vlined, linesnumbered]{algorithm2e}
\usepackage{graphicx}
\usepackage{textcomp}
\usepackage{xcolor}
\usepackage{listings}
\usepackage{tabu}
\usepackage{multirow}
\usepackage{adjustbox}
\usepackage{braket}
\usepackage{mathptmx}
\usepackage{amsfonts}
\usepackage{hyperref}
\usepackage{mathtools}
\usepackage{subcaption}
\usepackage{tikz}
\usepackage{comment}

\usepackage[inline]{enumitem}
\def\BibTeX{{\rm B\kern-.05em{\sc i\kern-.025em b}\kern-.08em
    T\kern-.1667em\lower.7ex\hbox{E}\kern-.125emX}}
\begin{document}

%\title{iCheck: An RDMA based Adaptive Application-Level Checkpointing System for HPC Applications}
%\title{Leveraging Resource-Aware Checkpointing for Fault Tolerance in Malleable MPI Applications}
\title{Designing an Adaptive Application-Level Checkpoint Management System for Malleable MPI Applications}

\author{\IEEEauthorblockN{Jophin John\IEEEauthorrefmark{1}, Michael Gerndt\IEEEauthorrefmark{2}}
\IEEEauthorblockA{\\Chair of Computer Architecture and Parallel Systems, Technische Universit{\"a}t M{\"u}nchen \\
Garching (near Munich), Germany} \\
Email: john@in.tum.de, gerndt@in.tum.de}
\maketitle
\begin{abstract}
Dynamic resource management opens up numerous opportunities in High Performance Computing. It improves the system-level services as well as application performance. Checkpointing can also be deemed as a system-level service and can reap the benefits offered by dynamism. A checkpointing system can have better resource availability by integrating with a malleable resource management system. In addition to fault tolerance, the checkpointing system can cater to the data redistribution demand of malleable applications during resource change. Therefore, we propose iCheck, an adaptive application-level checkpoint management system that can efficiently utilize the system and application level dynamism to provide better checkpointing and data redistribution services to applications.
\end{abstract}


\graphicspath{ {./images/} }
\section{Introduction}
\label{sec:intro}
Static resource management, still predominant in High-Performance Computing, takes away many optimization opportunities otherwise catered by dynamic resource management. This static behavior is visible from system services to applications~\cite{jsc,ahn2014flux}. The potential offered by dynamic applications using dynamic resources spans beyond better system utilization ~\cite{malleableapp,chadhaadaptive,john2020invasive,dynamicallocation}, and checkpointing is no exception~\cite{adaptivecp,adaptivecpcharm,jonasthesis}. Trends in dynamism research bring malleability\cite{perf_measurement, compres17,kale2002malleable,deepsea,admire,surveymalleability} into MPI\cite{mpi} and resource management infrastructure.

To yield the benefits of such a malleable infrastructure, the checkpointing system must be able to react to the dynamic needs of resource managers and applications by adapting itself to the changing system requirements. This can result in faster checkpoint transfers and efficient utilization of checkpointable resources (available memory, parallel file system bandwidth). Furthermore, a checkpointing system can conduct data redistribution during resource change, thereby addressing one of the key challenges in writing malleable applications. 

Hence, we posit checkpointing systems as a resource and data management system that can simultaneously handle the nuances in checkpointing requirements of multiple applications and interact with resource managers to provide seamless checkpointing capabilities. Towards this, we propose iCheck, a multi-level adaptive checkpoint management system that can reconfigure its checkpointing resources on the fly and offers application-level checkpointing and data redistribution services to malleable MPI applications. This approach of application-level checkpointing as an adaptable resource management and data distribution service brings novel contributions to the checkpointing domain. %w.r.t. state-of-the-art checkpointing systems.

% The key contributions  in this work are:
% \begin{itemize}
%   \item A malleable multilevel application-level Checkpoint/Restart library \verb|iCheck| that supports data redistribution during resource change.
%   \item SLURM scheduler plugin that focusses on checkpointing
%   \item Support for malleable MPI implementation
% \end{itemize}
%\input{sections/related_work}
%\input{sections/background}
\section{iCheck design}
\label{sec:design}
\begin{figure}[ht]
    \centering
    \includegraphics[width=0.45\textwidth]{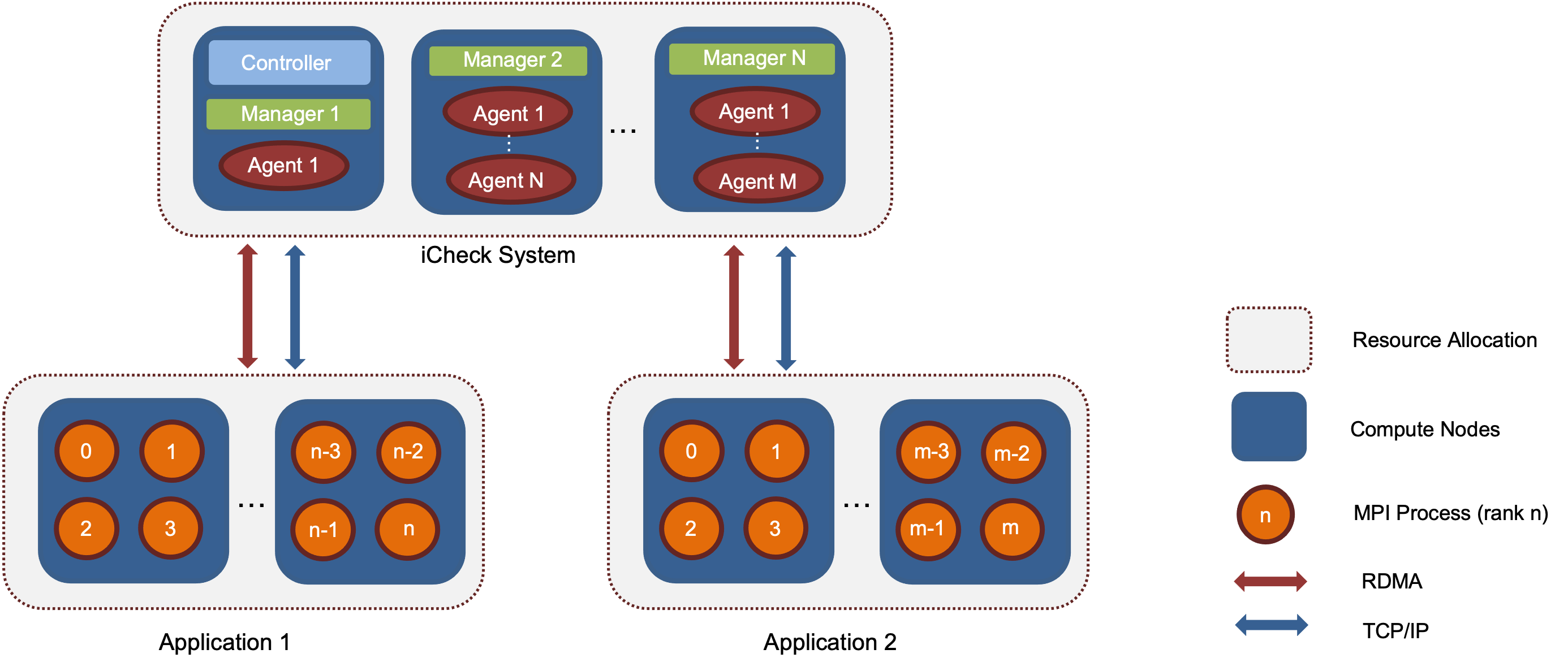}
    \caption{Components in iCheck system}
    \label{fig:ic_arch_comp}
\end{figure}

%iCheck is a checkpoint management system that provides checkpointing and data redistribution services to applications, orchestrates checkpoint transfers to the parallel file system,  monitors the checkpointable resources, and interacts with the resource manager to request/give back the checkpointable resources. 
iCheck has a core system and a library. iCheck core runs on a dynamically configurable number of dedicated nodes (iCheck nodes) of the HPC system and communicates with connected applications using the iCheck library (See Figure~\ref{fig:ic_arch_comp}). Application checkpoints are transferred to the memory of iCheck nodes using Remote Direct Memory Access (RDMA) and later written into the Parallel File System (PFS). The core system consists of an Agent, Manager, and Controller. The agent performs the functionality of checkpoint read/write (using libfabric\cite{fabriclibrary}) and data redistribution (for malleable implementations). Multiple agents can be assigned to a single application, and iCheck can dynamically change the agent count to obtain an optimum checkpoint transfer rate. The manager manages the node-level activities of the software, such as launching the agents and monitoring and predicting the node usage parameters (e.g., memory usage, bandwidth usage). The controller has a global view and performs the agent and node selection for connected applications based on the iCheck agent scheduling policies. These policies consider various system metrics (available memory, checkpoint frequency and size, and bandwidth usage) and can impact the overall checkpointing performance. The controller may also request the resource manager for additional resources based on resource availability. In addition, the controller will also orchestrate the writing of the checkpoint data into PFS by minimizing the effect on running applications.

%\subsection{Interactions in iCheck}
The iCheck core and iCheck library components in an application continually interact during the application life cycle. A generic workflow of these interactions is as follows:

During the start of the application:
\begin{enumerate}
\item An application registers itself with the controller.
\item The controller decides the number of agents and the iCheck nodes in which they will be launched.
\item The controller contacts the corresponding managers with the agent information.
\item The managers launch the agents in iCheck nodes and notify the controller.
\item The agents wait for the application processes to connect.
\item The controller provides the agent information to the application.
\item The application then connects with the agent.
\item The application and the agents register memory for RDMA.
\item The application and the agents perform checkpoint transfer operations.
\item Application finishes either with or without error.
\end{enumerate}
During the restart:
\begin{enumerate}
\item The application contacts the controller for checkpoint information.
\item The controller passes the agent information.
\item The application can either start by using the checkpoint from the agents or start new.
\end{enumerate}
During the data redistribution:
\begin{enumerate}
\item The application contacts the agent for data transfer.
\item The agents redistribute the data based on the mapping provided by the application.
\end{enumerate}

iCheck, with its agent-based checkpoint retrieval model, easily integrates the asynchronous checkpoint retrieval capability in its library. Since the agents use RDMA, the application does not need to block for data transfer rather it can continue the execution immediately after notifying the agents about the checkpoints. %The agents can remotely retrieve the data without significantly affecting the application performance.

\section{iCheck and Malleability}
\label{sec:iCheck}

\newcounter{mylabel}
\newcommand*\circled[1]{\refstepcounter{mylabel}\label{#1}\tikz[baseline=(char.base)]{
        \node[shape=circle,draw,inner sep=2pt] (char) {\arabic{mylabel}};}}

 %\circled{A}This is a  sample.  This sentence should \circled{B}
iCheck is malleable both at the system level and application level. At the system level, iCheck can horizontally scale its resources (available memory, number of agents) by attaching/removing iCheck nodes with the help of a resource manager. At the application level, iCheck can dynamically reconfigure the number of agents associated with connected applications.

\subsection{iCheck and Malleable Infrastructure} \label{metrics}
We have used the malleable infrastructure from the InvasiC project\cite{compres17,chadhaextend,john2020invasive} for our investigation into designing a malleable checkpointing system. It consists of an MPI implementation (an extension of MPICH with four additional methods) that can change the number of processes during the application execution and a malleable resource manager (extension of the Slurm \cite{yoo2003slurm}) that can orchestrate the resource changes in MPI. %Malleable MPI, an extension of MPICH, adds four additional methods to standard MPI to bring resource dynamism (See Listing \ref{icheckapi}). The malleable resource manager is an extension of the Slurm \cite{yoo2003slurm} to support resource-aware applications.

To support malleable MPI applications, the application-level checkpointing systems should be able to handle the changes in resources, i.e., whenever new processes are added, they should be able to reinitialize the system (most of the application-level checkpointing systems \cite{veloc,craft,fenix,scr,stagingnodes} needs to be modified). iCheck provide this capability out-of-the-box. Whenever a new set of processes are added to the application, iCheck can add new agents, thereby maintaining the checkpointing performance or reinitializing with the same agents. Additionally, using special APIs (\texttt{icheck\_add\_adapt()}, \texttt{icheck\_redistribute()} as in Listing \ref{icheckapi}), users can provide their data distribution mapping that can be used by iCheck when a resource change happens. iCheck also provides a special API call \texttt{icheck\_probe\_agents()} (line 29 in Listing \ref{icheckapi}) that can be used to query for agent change to improve the checkpointing time.

%\subsection{iCheck and Malleable Resource Manager}
To facilitate system-level malleability in iCheck, we have created an iCheck-aware job scheduling plugin in the resource manager (RM). The following interactions with iCheck are supported by the new plugin.\circled{1} RM can give resources (iCheck nodes) to the iCheck based on the request and availability. For example, when iCheck runs out of memory in a node, the controller can request more memory and get additional nodes from RM.\circled{2} RM can retake nodes from iCheck. For example, to support resource requirements of a priority job or to satisfy power requirements.\circled{3} RM can ask the controller to migrate resources to a different iCheck node.\circled{4} RM can pass application-specific information to the controller. For example, it can inform the controller about an impending resource change of an application so that agents can prepare for data redistribution ahead of time.

\subsection{iCheck library and a simple malleable MPI application}  \label{library}
The pseudocode of a naive iCheck enabled malleable MPI application is shown in Listing \ref{icheckapi}. The four malleable MPI routines behave as follows. \texttt{MPI\_Init\_adapt()} (line 3) will register the application with the resource manager and returns the type of the MPI process. A process can be initial (created as part of an initial application launch) or joining (created as part of an application expansion). Initial processes continue the application execution (skipping lines 10-15) and regularly call \texttt{MPI\_Probe\_adapt()} (line 17) to check for any resource change. On the other hand, joining processes immediately calls the  \texttt{MPI\_Comm\_adapt\_begin()} (line 11) collective function and waits for initial processes to join. Meanwhile, the \texttt{MPI\_Probe\_adapt()} (in line 17) call informs initial processes about the resource change triggered by the malleable resource manager. Initial processes, then calls the collective \texttt{MPI\_Comm\_adapt\_begin()} function (line 19). Once both the initial and joining processes call \texttt{MPI\_Comm\_adapt\_begin()}  routine (line 11 and line 19), they can perform data redistribution and share application-specific data and control information among them. After that, they call \texttt{MPI\_Comm\_adapt\_commit()} (lines 14 and 22) collective function to finish the resource adaptation and together resume the application execution. %The initial and joining processes will have the same characteristics and will continue the application execution.
\begin{minipage}{.48\textwidth}
 \lstset{language=C++, basicstyle=\small,captionpos=b,caption={Psuedocode of a malleable application with iCheck},label={icheckapi}, keepspaces=true, stepnumber=1, firstnumber=1, numberfirstline=true, 
                 keywordstyle=\color{blue}\ttfamily,
                 stringstyle=\color{red}\ttfamily,
                 commentstyle=\color{gray}\ttfamily,
                 numbers=left,
                 morecomment=[l][\color{magenta}]{\#},
                 xleftmargin=2.6em,
                 framexleftmargin=2.7em
                 }
 \begin{lstlisting}[frame=single][belowskip=-0.8 \baselineskip]
#include<icheck.h>
int main() {
 MPI_Init_adapt(..., type)
 float data[SIZE];
 icheck_init(..., type);
 icheck_add_adapt("data",data,...,BLOCK);
 if(checkpoint_available && no_adapt){
   icheck_restart();
 }
 if (type == joining) {
     MPI_Comm_adapt_begin(...);
     icheck_redistribute("data",data, 
                new_size, BLOCK)
     MPI_Comm_adapt_commit();
 }
 while (true){
    MPI_Probe_adapt(resource_change,...);
    if (resource_change) {
        MPI_Comm_adapt_begin(...);
        icheck_redistribute("data",data, 
                new_size, BLOCK)
        MPI_Comm_adapt_commit();
    }
    /*Read/Write data[]*/
    if(iteration%100)
        icheck_commit();
    /*Check for agent change*/
    if(iteration%1000)
        icheck_probe_agents();
    }
    icheck_finalize(...);
    MPI_Finalize();
}
 \end{lstlisting}
 \end{minipage}\hfill
 
\texttt{icheck\_init()} (line 5) registers the application with the controller. \texttt{icheck\_add\_adapt()} (line 6) is used to add the data to be checkpointed along with the mapping needed for data redistribution during a resource adaptation. If present, the initial processes can restart and restore the previously stored checkpoint \texttt{icheck\_restart()} (lines 7-9). Otherwise, it continues the application execution. \texttt{icheck\_commit()} (in line 26) will trigger agents to perform asynchronous checkpoint transfer to iCheck nodes. During an application expansion, initial and joining processes will call the \texttt{icheck\_redistribute()} (line 12 and line 20) to get the proper data to resume the application execution. Currently, iCheck only supports naive data redistribution schemes like block (line 12 and line 20) or cyclic during a resource change.
\section{related work}
\label{sec:rw}
There are a plethora of checkpointing works based on different categories like application-level or system-level, process recovery or data recovery, multilevel or single-level, coordinated or uncoordinated, and blocking or non-blocking \cite{ulfm,rmpi,ftccharm,ftmpi,cppc,ereinit,mpistages,porch,noninvasive_alc,italc,libckpt,mcrengine,burstbuffer,in_system_burst_buffer,stagingnodes,veloc,craft,fenix,scr}. Here we focus on works closest to iCheck.

iCheck is an adaptive asynchronous multilevel application-level in-memory checkpointing system using RDMA that provides data distribution service and fault tolerance. To the best of our knowledge, no single system exists with all the above characteristics for application-level checkpointing. Existing application-level checkpointing libraries (for example, \cite{fenix, craft, scr}) cannot reconfigure their resources across multiple applications to optimize their checkpointing activity. In contrast, iCheck can centrally manage checkpoints and improve the overall checkpointing performance by dynamically scaling its resources.

Regarding multilevel RDMA capability in checkpointing, the closest work to iCheck is by Sato et al.\cite{stagingnodes}. iCheck differs in the following aspects. iCheck can simultaneously cater to the needs of multiple applications and reconfigure its resources for RDMA capability on the fly. Further, iCheck does not use extra processes inside the compute nodes of applications to transfer the checkpoint.

VeloC \cite{veloc} is a related work to iCheck with regards to the adaptivity in checkpointing. It is an adaptive asynchronous checkpointing system where adaptivity refers to efficiently selecting the underlying checkpoint storage based on the available heterogeneous solutions. In iCheck, adaptivity refers to the flexibility in scaling the resources of our system.

%\input{sections/evaluation}
%\section{Holistic discussion of results and outlook towards future}
%\section{Challenges and Open Questions}
%First, the authors need to motivate the decoupled design they adopted better. Specifically, checkpoints are captured on remote dedicated nodes, which act as a checkpointing service where multi-level checkpointing techniques are applied. However, transferring checkpoints to remote nodes blocks the application for a long time at each checkpoint request, which generates high checkpointing overhead in addition to the core-hours consumed by the dedicated checkpointing nodes. Instead, asynchronous multi-level checkpointing techniques have been successfully applied directly on the compute nodes (e.g. write checkpoints to local storage in blocking fashion, flush to a parallel file system in the background), thereby reducing the checkpointing overhead without consuming extra core-hours.
%In particular, what are the challenges of making checkpointing solutions malleable? How do we decide when to add/remove nodes/resoures allocated to checkpointing? How do we choose whether to prioritize checkpointing overhead vs. core-hours? Many multi-level checkpointing techniques introduce redundancy that is distributed across multiple nodes (e.g. replicas, erasure codes, etc.). If we remove checkpointing nodes/resources, how do we redistribute the redundancy efficiently among fewer nodes to guarantee the same resilience properties?
\section{conclusion and future work}
\label{sec:conc}

%The experimental iCheck system is working in a virtual cluster of 8 nodes. 
%iCheck is a self-adaptive checkpointing system that changes its resources based on the characteristics of running applications. It provides checkpointing and data redistribution services to applications. iCheck centrally manages checkpoint resources to ensure better-checkpointing services to all the running applications.

Our current work on iCheck focuses on analyzing the malleable characteristics of complex applications and adapting iCheck to support the intricacies of data distribution in such applications. The current naive data redistribution schemes in iCheck are insufficient to utilize malleability in complex applications (for example, data redistribution inside molecular dynamics applications during a resource change).

Furthermore, we intend to extend the current iCheck-aware job scheduling plugin inside RM. The current experimental plugin prioritizes iCheck and will allocate nodes to iCheck based on the controller's request and node availability. We will extend it by incorporating a malleable resource management policy that removes nodes from running applications based on their performance~\cite{chadhaextend}, ensuring fairness to running jobs and jobs waiting in the queue.

Our prototype implementation is currently deployed on a virtual cluster and analyzed with synthetic malleable applications. We plan to evaluate iCheck on SuperMUC-NG\cite{lrz_supermuc} with real scientific applications developed using malleable MPI.

\section{Acknowledgment}
\label{sec:ack}
% The research leading to these results has received funding from the European Union’s Horizon 2020 Programme under grant agreement number 671657. We thank the Centre for Information Services and High Performance Computing (ZIH) at TU Dresden for providing HPC resources that contributed to our research.
The research leading to these results was funded by the Deutsche Forschungsgemeinschaft 
(DFG, German Research Foundation)-Projektnummer 146371743-TRR 89: Invasive Computing.
%This document is a model and instructions for \LaTeX.
%Please observe the conference page limits. \cite{chadha_upapi}

%\section*{Acknowledgment}
%The research leading to these results was funded by the Deutsche Forschungsgemeinschaft 
%(DFG, German Research Foundation)-Projektnummer 146371743-TRR 89: Invasive Computing.

\bibliographystyle{IEEEtran}
\bibliography{reference}

\end{document}